\documentclass[10pt,conference]{IEEEtran}
\IEEEoverridecommandlockouts
\usepackage{cite}
\usepackage{amsmath,amssymb,amsfonts}
\usepackage{algorithmic}
\usepackage{graphicx}
\usepackage{textcomp}
\usepackage{xcolor}
\usepackage{url}
\usepackage{pgfplots}
\pgfplotsset{compat=1.18}
\usepackage{tikz}
\usepackage{adjustbox}
\usepackage{soul}
\usepackage{tcolorbox}
\usepackage{booktabs}

\usepackage{algorithm}
\usepackage{algorithmic}

\usepackage{listings}
\usepackage{varwidth} % utile per minipage flessibile
\newsavebox{\jsonbox} % buffer per contenuto verbatim

\lstdefinestyle{jsonstyle}{
  backgroundcolor=\color{gray!10}, 
  basicstyle=\scriptsize\ttfamily,
  frame=single,
  breaklines=true,
  postbreak=\mbox{\textcolor{gray}{$\hookrightarrow$}\space},
  showstringspaces=false,
  captionpos=b,
  language=,
}

\def\BibTeX{{\rm B\kern-.05em{\sc i\kern-.025em b}\kern-.08em
    T\kern-.1667em\lower.7ex\hbox{E}\kern-.125emX}}

\usepackage{etoolbox}

\begin{document}

\title{A forgery attack on the Block.co blockchain-based digital credential certification system
%\thanks{This work was partially supported by project SERICS (PE00000014) under the Italian Ministry of University and Research (MUR) National Recovery and Resilience Plan, funded by the European Union - Next Generation EU.}
}

 \author{
  \IEEEauthorblockN{Giacomo Zonneveld, Giulia Rafaiani and Marco Baldi}
  \IEEEauthorblockA{ Department of Information Engineering, Università Politecnica delle Marche, Ancona, Italy \\ Email: \{g.zonneveld, g.rafaiani, m.baldi\}@univpm.it}
  }

\maketitle

\begin{abstract}
Certification of digital documents, such as academic credentials, seems a particularly suitable application for the use of blockchain and distributed ledger technologies. Indeed, these technologies enable decentralized certification systems that rely on the immutability and persistence of their distributed ledgers. However, in the absence of a central trusted authority, it is not easy to guarantee the authenticity of the connection between the real identity of an academic institution and the digital identity of the certificate issuer. In this paper, we demonstrate that one of such systems, known as Block.co, has a vulnerability that allows the production of forged certificates that are recognized as valid by the system. Since this is an inherent limitation of the approach used for blockchain-based certification, our attack is likely to be extendable to other systems adopting the same approach.
\end{abstract}


\section{Introduction}

The emergence of blockchain and distributed ledger technology (DLT) has introduced new opportunities for data management and certification, particularly in scenarios that require high levels of security and trust.
In fact, these decentralized technologies provide public ledgers that are time-stamped and immutable, supporting the sequential storage of digital transactions and other data in an immutable and publicly verifiable way.
These characteristics lead to a typical use of these technologies for security purposes, which is sometimes named \textit{proof of existence} of generic digital objects. In a nutshell, it is accomplished by crystallizing onto the public ledger, within a specific transaction, a cryptographic proof (e.g., a hash digest or a Merkle proof) of the existence of some digital artifact at a certain time, which is then associated with the transaction identifier that proves its existence and integrity at that time.
A relevant application of this approach is within the education sector, where systems for issuing academic and professional certificates can exploit decentralized ledgers to ensure their integrity, authenticity and availability over time, without the need for central authorities. Indeed, using blockchain and DLT for certifying academic credentials can guarantee their integrity and their verifiability everywhere and over time, even after that the credential issuer disappears.

However, the adoption of decentralized technologies can also bring with it some specific security issues. In fact, although the immutability and decentralization of these digital infrastructures provide strong guarantees against data tampering, vulnerabilities can still emerge at different levels, particularly with regard to verifying the identity of the subject issuing the credentials.
In fact, the purely decentralized nature of blockchain and DLTs significantly limits the role of any trusted authority or governing body, making the connection between users and their real-world identities somehow loose, even when explicitly declared by user themselves.

This unfortunately makes it easier to mount attacks aimed at impersonating the issuer of academic credentials and forging the credentials themselves.
In fact, if we want to preserve the decentralized nature of the system, the real identity of the credential issuer cannot be certified through centralized techniques, as is normally done with a public key infrastructure (PKI) and Certification Authorities (CAs), but must rather be declared in some way by the issuer itself. This does not prevent a different party from claiming to be the real issuer of the credentials, and generating forged credentials.
Moreover, the user who is the legitimate issuer is not notified in any way that others are claiming to be the same.

A solution that would seem to be applicable in this context is represented by World Wide Web Consortium (W3C) standards for Decentralized Identifiers (DIDs) and Verifiable Credentials (VCs), which are key features towards Self-Sovereign Identity (SSI) and aim at reducing reliance on centralized nodes for managing identities. However, these solutions do not eliminate the need for trusted authorities. Indeed, decentralization of identifiers ensures their immutability, availability and integrity from a technical standpoint; however, the authenticity of attributes is still based on the use of trust anchors and authorities, which are inherently centralized.

The attack we present is inspired by a previous attack targeting a blockchain-based academic credential issuing and certification system based on Blockcerts \cite{Baldi2020}.
In this paper, we show that an attack of a similar type can also be mounted against another academic credential issuing and certification system called Block.co.
Although the impersonation attack we propose is based on the same theoretical principles as the one described in \cite{Baldi2020}, there are substantial differences in its design and attack vector. First, the credential format differs: Blockcerts operates on JSON artifacts compliant with the Open Badges standard, while the system analyzed in this work integrates cryptographic evidence directly into the metadata of PDF documents. Second, the used verification/validation logic is different. While Blockcerts is natively decentralized, Block.co adopts a hybrid architecture. Therefore, the attack requires, as an additional step, manipulation of a specific field in order to inhibit centralized registry control and force the validator to use a decentralized mechanism. 
Table \ref{tab:novelty} summarizes the main differences between the two attacks.

\begin{table}[th]
\centering
\caption{Comparison with the Blockcerts forgery attack \cite{Baldi2020}}
\label{tab:novelty}
\scriptsize
\begin{tabular}{l|l|l}
\toprule
& \textbf{Blockcerts \cite{Baldi2020}} & \textbf{Block.co (this work)} \\
\midrule
Credential format & JSON (Open Badges) & PDF + metadata \\
\hline
Issuer binding & Hosted JSON profile & URL + Bitcoin wallet \\
\hline
Cryptography & Digital signature & Merkle anchoring only \\
\hline
Architecture & Fully decentralized & Hybrid (centr. + decentr.) \\
\hline
Bypass needed & No & Yes (\textit{block\_co=false} required) \\
\hline
Attack artifacts & Fake JSON + key pair & Fake URL + Bitcoin wallet \\
\bottomrule
\end{tabular}
\end{table}

In the following, we first describe how the Block.co system is designed, showing why the connection between real identities and system users is loose. 
Then, we describe the structure of the digital credentials it generates, and how the issuer could be impersonated by a malicious user.
Finally, we validate our approach through a practical example, by constructing a forged certificate and showing how the system erroneously accepts it as valid.
The paper is organized as follows.
Section \ref{sec:relatedworks} describes related works and Section \ref{sec:background} introduces the necessary background. 
Section \ref{sec:issuance} describes blockchain-based methods for digital credentials issuance, with focus on the Block.co approach.
In Section \ref{sec:forgery} we illustrate how to mount a forgery attack on the Block.co protocol and in Section \ref{sec:countermeasures} we propose some possible countermeasures.
In Section \ref{sec:disclosure} we report details about the responsible disclosure of the attack, and Section \ref{sec:concl} provides some concluding remarks.

\section{Related works} \label{sec:relatedworks}
Conventional solutions for certification of digital credentials rely on PKI mechanisms, whereby an academic institution generates a pair of public and private keys and obtains a public key certificate from a CA. The issued credentials are then certified through the application of a digital signature using the institution's private key, so that the authenticity and integrity of the credentials are validated through the institution's public certificate \cite{Boonkrong2024}. Although strong cryptographic guarantees are provided, these approaches suffer from limitations relating to the integration of a CA and the availability of the institution's digital certificate. 
They are, in fact, centralized and require a chain of trust that must be maintained over time so that credentials can be verified even long after they were issued.

To avoid having to resort to PKIs and certification authorities, a basic solution such as the one proposed in \cite{Boonkrong2024} could be considered, which consists in generating the hash digest of the credential, then storing the digest and the credential itself in a database maintained by the issuer. Although being a straightforward solution, it has many limitations due to centralization: the database is a single point of failure and once an issuer disappears, the emitted credentials are no longer verifiable.
Such limitations due to centralization can be overcome by using blockchain technology, and saving digital fingerprints of the credentials on the blockchain through a transaction \cite{blockcerts}. In \cite{btcert} the authenticity of the credentials issuer is provided through digital signatures, which, however, require the distribution of the corresponding public keys authenticated by some central authority.
In \cite{eductx} only accredited institutions, acting as validating nodes, can produce and distribute credentials. Each credential is encrypted with the credential holder public key and stored on InterPlanetary File System (IPFS). The resulting IPFS content identified (CID) together with the hash of the credential is sent to the address of the recipient through a transaction. The system’s permissioned design limits openness and scalability \cite{fakeme}, as well as verifiability of credentials in the event that the platform goes offline.
In \cite{Ghazali}, instead, the blockchain is used in combination with digital signatures. 
As stated in \cite{fakeme}, however, both \cite{eductx} and \cite{Ghazali} suffer from an issue involving unauthorized authorities that can easily register and produce fake certificates. 

Other blockchain-based solutions leverage smart contracts. In \cite{Gresch_2018} certificate hashes are saved in a smart contract owned by the issuing authority. In \cite{Serranito2020} a more complex solution is proposed: authorities must participate in a consortium regulated through a smart contract which manages authorities identification (through DIDs), admission through voting and governance. Once admitted, each authority can certify credentials by publishing hashes in a owned smart contract as in \cite{Gresch_2018}.
As stated in \cite{fakeme}, while the use of DIDs and Zero-Knowledge Proofs (ZKPs) \cite{ZKP1}\cite{ZKP2} may guarantee enhanced privacy and user autonomy, they also come with limitations related to the absence of common standards.
Above all, the main limitation remains, that is, the fact that in the absence of some central authority the issuer's authenticity cannot be guaranteed.

\section{Background}
\label{sec:background}

%\subsection{Blockchain-based credential anchoring}

Digital credentials can be anchored to a blockchain to establish tamper-proof timestamped records, and the use of Merkle trees enable efficient aggregation of multiple credential hashes into a single blockchain transaction. 
This approach, often called \textit{proof of existence}, crystallizes a  cryptographic fingerprint of a digital artifact onto a public ledger  within a specific transaction, associating it with a transaction identifier that proves its existence and integrity at a particular time. The certification proof of such artifacts can be provided in a standardized format called \textit{Chainpoint Proof} \cite{chainpoint}, which  contains all the information required for verifying their integrity.

\subsection{Merkle trees and proof verification}

Merkle trees are binary tree structures built from a set of data divided 
into blocks, designed for efficient use of hash functions in data integrity 
verification. The hash digest of each data block forms a leaf of the tree. 
Internal nodes are progressively constructed by computing the hash of 
adjacent pairs of child nodes, recursively, until reaching the root, a 
single hash digest known as the \textit{Merkle root}.
Key properties of Merkle trees are:
\begin{itemize}
    \item \textbf{Integrity guarantee:} Any modification to data in a leaf 
    results in a change of the Merkle root.
    \item \textbf{Efficiency:} To verify a specific piece of data, only the 
    Merkle root and the list of hash digests connecting the data block to the root, namely the \textit{Merkle proof}, are needed. The computation and storage costs are both $\mathcal{O}(\log_2(N))$.
    \item \textbf{Privacy:} Merkle proof verification does not require knowledge of data corresponding to the other leaves.
\end{itemize}

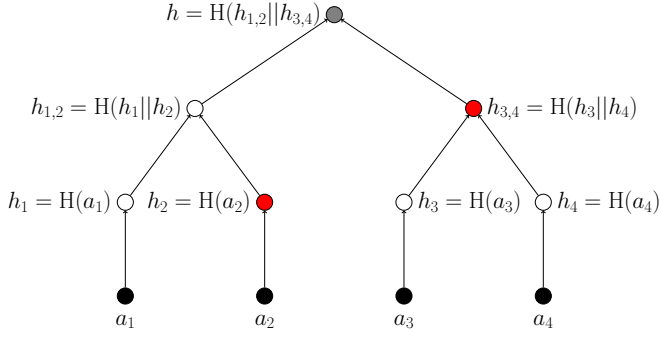
\begin{figure}
    \centering
    \resizebox{\columnwidth}{!}{
\begin{tikzpicture}

\newcommand{\hdist}{2.6}
\newcommand{\vdist}{3.5}

\begin{scope}[every node/.style={fill = black, circle,thick,draw}, minimum size = 0.6cm]
\node[label = south:{\Huge $a_1$}] (V01) at (-3*\hdist,0) {};
\node[label = south:{\Huge $a_2$}] (V02) at (-\hdist,0) {};
\node[label = south:{\Huge $a_3$}] (V03) at (\hdist,0) {};
\node[label = south:{\Huge $a_4$}] (V04) at (3*\hdist,0) {};
\end{scope}

\begin{scope}[every node/.style={ circle,thick,draw}, minimum size = 0.6cm]
\node[label = west:{\Huge $h_1 = \mathrm{H}(a_1)$}] (H01) at (-3*\hdist,\vdist) {};
\node[fill=red, label = west:{\Huge $h_2=\mathrm{H}(a_2)$}] (H02) at (-\hdist,\vdist) {};
\node[label = east:{\Huge $h_3= \mathrm{H}(a_3)$}] (H03) at (\hdist,\vdist) {};
\node[label = east:{\Huge $h_4= \mathrm{H}(a_4)$}] (H04) at (3*\hdist,\vdist) {};
\end{scope}

\begin{scope}[every node/.style={ circle,thick,draw}, minimum size = 0.6cm]
\node[label = west:{\Huge $h_{1,2} = \mathrm{H}(h_1|| h_2)$}] (H11) at (-2*\hdist,2*\vdist) {};
\node[fill=red, label = east:{\Huge $h_{3,4} = \mathrm{H}(h_3|| h_4)$}] (H12) at (2*\hdist,2*\vdist) {};
\end{scope}

\begin{scope}[every node/.style={ circle, fill = gray,thick,draw}, minimum size = 0.6cm]
\node[label = west:{\Huge $h = \mathrm{H}(h_{1,2}|| h_{3,4})$}] (H21) at (0,3*\vdist) {};
\end{scope}

\path [->, thick] (V01) edge (H01);
\path [->, thick] (V02) edge (H02);
\path [->, thick] (V03) edge (H03);
\path [->, thick] (V04) edge (H04);

\path [->, thick] (H01) edge (H11);
\path [->, thick] (H02) edge (H11);

\path [->, thick] (H03) edge (H12);
\path [->, thick] (H04) edge (H12);

\path [->, thick] (H11) edge (H21);
\path [->, thick] (H12) edge (H21);

\end{tikzpicture}
}
    \caption{Example of a Merkle tree with 8 leaves. $\mathrm{H}$ represents 
    the hash function. The digest $h$ (gray node) is the Merkle root, while 
    the digests associated to the nodes in red form the Merkle proof for 
    the node $a_1$.}
    \label{fig:mt}
\end{figure}

An example of a Merkle tree with 8 leaves is shown in Fig. \ref{fig:mt}, 
with the Merkle proof for leaf $a_1$ highlighted in red.

\subsection{Chainpoint}

Chainpoint \cite{chainpoint} is an open standard for blockchain-based certification of multiple data of any type. In particular, the existence and integrity of multiple data can be certified through a single blockchain transaction, by leveraging Merkle trees.
The certification proof is provided in a standardized format called \textit{Chainpoint Proof}, which contains all the information required for verifying the integrity of data. 
There are currently three versions of the Chainpoint Proof format, but we focus on the one used by Block.co, which is version 2.0.
A version 2.0 Chainpoint Proof, of which we provide an example in Fig. \ref{fig:chainpoint}, includes the following fields:
\begin{itemize}
    \item \textbf{@context} - specifies the JSON-LD context for the proof.
    \item \textbf{type} - specifies the hash method and the \textit{Chainpoint Proof} version.
    \item \textbf{targetHash} - contains the hash digest value of the certified data.
    \item \textbf{merkleRoot} - contains the Merkle root stored into the blockchain transaction.
    \item \textbf{proof} - contains the Merkle proof for the certified data.
    \item \textbf{anchors} - contains some blockchain-related information depending on the adopted blockchain (Ethereum or Bitcoin), like the transaction field used to store the Merkle root and the transaction identifier.
\end{itemize}

\begin{figure}[ht]
\centering
{\scriptsize
\begin{lrbox}{\jsonbox}
\begin{minipage}{.95\linewidth}
\begin{verbatim}
{
  "@context": "https://w3id.org/chainpoint/v2",
  "type": "ChainpointSHA256v2",
  "targetHash": "bdf8c9bdf076d6aff0292a1c9448691d2ae28
                    3f2ce41b045355e2c8cb8e85ef2",
  "merkleRoot": "51296468ea48ddbcc546abb85b935c73058fd
                    8acdb0b953da6aa1ae966581a7a",
  "proof": [
        {"left": "bdf8c9bdf076d6aff0292a1c9448691d2ae28
                    3f2ce41b045355e2c8cb8e85ef2"},
        {"left": "cb0dbbedb5ec5363e39be9fc43f56f321e157
                    2cfcf304d26fc67cb6ea2e49faf"},
        {"right": "cb0dbbedb5ec5363e39be9fc43f56f321e15
                    72cfcf304d26fc67cb6ea2e49faf"}
  ],
  "anchors": [
    {"type": "BTCOpReturn",
      "sourceId": "f3be82fe1b5d8f18e009cb9a491781289d2e0
                    1678311fe2b2e4e84381aafadee"}
  ]
}
\end{verbatim}
\end{minipage}
\end{lrbox}

\fcolorbox{gray}{gray!10}{\usebox{\jsonbox}}
}
\caption{Example of a version 2.0 Chainpoint Proof from \cite{chainpoint}.}
\label{fig:chainpoint}
\end{figure}

\section{Blockchain-certified digital credentials
\label{sec:issuance}
}
There are several examples in the literature of decentralized systems based on blockchain and DLT for the issuance and certification of digital credentials.
A straightforward approach to this aim is to compute a cryptographic hash digest of some digital certificate, then storing the digest into a public blockchain (e.g., Bitcoin, Ethereum). The actual certificate (PDF/JSON), instead, can be stored off-chain, for example using IPFS (InterPlanetary File System) or any other physical or cloud-based data storage facility.
When using this approach, the verification of the digital credential is simply performed by recomputing its hash digest and comparing it with the one stored in the blockchain.
If the two digests match, then the integrity of the credential is verified.
This approach is the one used, among others, by the Blockcerts protocol\cite{blockcerts}, which indeed was subject to the forgery attack presented in \cite{Baldi2020}.

The European Blockchain Service Infrastructure (EBSI) proposes a different approach for issuing digital certificates \cite{ebsi}. In this case, institutions issue W3C-compliant VCs. Blockchain is used to register issuer identifiers, timestamps, or revocation registries, but the connection with real identities is still guaranteed by trusted authorities. %\textcolor{red}{è vero?}

Other approaches exploit, for example, permissioned blockchains \cite{permissioned} or NFTs (Non-Fungible Tokens) and tokenization in general \cite{nft}, as well as smart contracts that are capable to manage the creation, revocation, and validation of the certificates \cite{sc}.

\subsection{Block.co protocol for certification of digital credentials}

Block.co is a service providing issuance of credentials as a proof of some kind of achievement accomplished by someone \cite{blockco}. 
Each digital credential is produced in the form of a PDF certificate, and the hash fingerprint of the certificate is stored on the Bitcoin blockchain as a guarantee of authenticity and prevention of forgery. As a design principle, the service proposes a decentralized solution to tackle the dependency on the issuer for the verification of the digital credential. In this way, it should always be possible to verify a credential, even if the issuer no longer exists or has lost the corresponding verification record \cite{blockcodesign}. 

The Block.co solution is aimed at allowing institutions to issue digital credentials representing certificates, diplomas, driving licenses, etc. The platform provides the certified credential as a PDF, with information concerning its blockchain-based certification embedded within its metadata.

\subsubsection{Certificate structure}

Each Block.co certificate file contains all the information related to the correponding achievement and to its validity, such as the certificate itself, its issuer name, its recipient name and its blockchain-based integrity proof. For this purpose, ad-hoc PDF metadata are inserted into the PDF file, such as:
\begin{itemize}
    \item \textbf{Issuer}: information about the credential issuer.
    \item \textbf{Metadata}: information about the certificate owner.
    \item \textbf{Chainpoint Proof}: information about the blockchain and the integrity proof, according to the Chainpoint standard.
    \item \textbf{Version}: version of the metadata format adopted.
\end{itemize}

\subsubsection{Issuance procedure}

%Supposing the issuer produces multiple certificates for different attendees obtaining the same achievement, multiple PDFs are created.

When some credential needs to be certified, first of all, a special PDF file for each credential is generated. 
Besides the main contents describing the credential, some metadata are inserted into the created file \cite{blockcopaper}, including the \textit{Issuer} field, the \textit{Metadata} field and an empty \textit{Chainpoint Proof} field. Then, a group of prepared files of this type are used as leaf nodes to build a Merkle tree and obtain the corresponding Merkle root. 
Such a Merkle root, together with a prefix identifying the issuer, is then written into the Bitcoin blockchain inside the \textit{OP\_RETURN} field of a Bitcoin transaction that is stored into the Bitcoin ledger. 
After the transaction has been accepted by the Bitcoin network, its transaction identifier and the Merkle proof are used to fill the \textit{Chainpoint Proof} field of each certified credential PDF file. % as shown in Fig. \ref{fig:chainpoint}.

%\begin{figure}[htbp]
%\centering
%{\scriptsize
%\begin{minipage}{0.95\linewidth}
%\begin{verbatim}
%{
%  "@context": "https://w3id.org/chainpoint/v2",
%  "type": "ChainpointSHA256v2",
%  "targetHash": "leaf hash of the single certificate",
%  "merkleRoot": "merkle root of the tree",
%  "proof": "merkle proof",
%  "anchors": [
%    {
%      "type": "BTCOpReturn",
%      "sourceId": "transaction hash"
%    }
%  ]
%}
%\end{verbatim}
%\end{minipage}
%}
%\caption{Chainpoint Proof format}
%\label{fig:chainpoint}
%\end{figure}

At this point, the PDF is certified and ready to be shared and verified.

\subsubsection{Certificate verification procedure}

The validation of a certified PDF file starts with the extraction of the \textit{Chainpoint Proof} metadata from the certified PDF file, from which important information is retrieved, namely, the transaction identifier and the Merkle proof. 
The \textit{Issuer} metadata is also read, but it is not removed from the file, differently from the \textit{Chainpoint Proof} metadata, since it is part of the certified information. 
Its structure is shown in Fig. \ref{fig:std-issuer}. 

\begin{figure}[th]

\centering
{\scriptsize
\begin{lrbox}{\jsonbox}
\begin{minipage}{0.9\linewidth}

\begin{verbatim}
{ 
    "name": "Name of the issuer, i.e. Univpm",
    "identity": {
        "address": "Bitcoin address",
        "verification":[
            {"block_co": "true/false"}, 
            {"domain": "URL of the issuer website"}
        ]
    }
}
\end{verbatim}

\end{minipage}
\end{lrbox}

\fcolorbox{gray}{gray!10}{\usebox{\jsonbox}}
}
\caption{Issuer metadata format}
\label{fig:std-issuer}
\end{figure}

Based on the value of the \textit{block\_co} flag, two types of identity verification are possible for the issuer:
\begin{itemize}
    \item \textit{block\_co = true}: \textbf{internally} to the platform, based on a centralized list;
    \item \textit{block\_co = false}: \textbf{externally} to it, by comparing the address with the one published by the issuer itself in a hosted text file; this file should be accessible through the URL published within the \textit{domain} field.
\end{itemize}
These two mechanisms can jointly be used on the same document, thus achieving both centralized and decentralized issuer authentication. However, centralized authentication depends on the robustness of the centralized list, and is likely to be weaker than what a trusted identity provider and a PKI can provide, without the need for blockchain technology. 
Decentralized issuer authentication, on the other hand, relies on blockchain technology but is vulnerable to forgery attacks, as we will show next.
If the \textit{block\_co} field is omitted, its default value is assumed to be false.

The \textit{address} field is used to retrieve from the blockchain all the transactions originated by that address. 
The transaction identifier written within the \textit{Chainpoint Proof} metadata is then searched in such a list of transactions.
The blockchain transaction containing the Merkle root corresponding to the certified file is so retrieved.
Then, the PDF file is hashed and its digest is used along with the Merkle proof included in the metadata to compute a local copy of the Merkle root. 
At this point, the original value of the Merkle root computed when the credentials were issued is retrieved from the OP\_RETURN field of the identified blockchain transaction and compared with the locally computed one.
If they match, the credentials are considered valid, otherwise the file is considered altered and no longer reliable \cite{blockcopaper}. 

%Also a verification on possible credentials revocation is done but it is out of scope. 

%The commercialized platform also performs some verification on the issuer. The \textit{Issuer} metadata is structured as shown in Fig. \ref{fig:std-issuer}.
%\begin{figure}[htbp]
%\centering
%{\scriptsize
%\begin{minipage}{0.95\linewidth}
%\begin{verbatim}
%{ "name": "Name of the issuer, i.e. Univpm",
%    "identity": {
%        "address": ``Bitcoin address",
%        "verification":[
%            {"block_co": "true/false"}, 
%            {"domain": "URL of the issuer website"}
%        ]
%    }
%}
%\end{verbatim}
%\end{minipage}
%}
%\caption{Issuer metadata format}
%\label{fig:std-issuer}
%\end{figure}

%By performing some tests, we find that the \textit{address} field is used to retrieve all the transactions from which the one present in the \textit{Chainpoint Proof} is searched. The \textit{verification} field is used instead for the issuer validation. If the \textit{block\_co} field is set to \textit{true}, the address \textcolor{red}{authorization} is checked internally to the platform. The \textit{domain} field is used to search for a published address on the endpoint that matches the one found in the \st{\textit{Chainpoint Proof}} \textcolor{red}{\textit{Issuer} metadata o transaction sender address o entrambi}. 

\begin{figure}[th]
\centering
\includegraphics[width=0.95\linewidth]{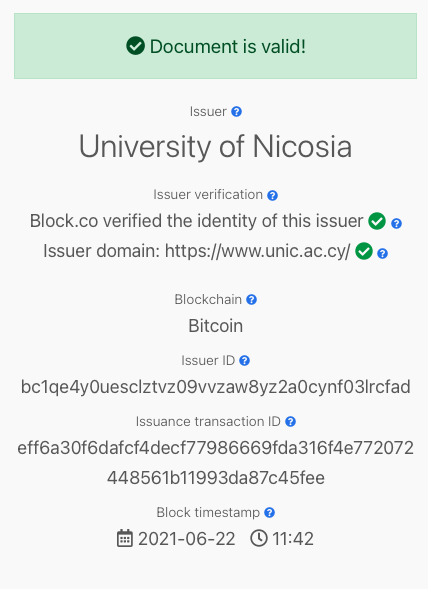}
\caption{Verification result of an authentic PDF}
\label{fig:verification-normal}
\end{figure}

Fig. \ref{fig:verification-normal} shows the outcome of the verification of an authentic credential file. 
As we see from the figure, the information reported upon successful verification is the overall verification result, the result of the issuer identity verification and some data concerning the Bitcoin transaction.
%\textcolor{red}{Aggiungerei due parole per spiegare che in questo caso la verifica dell'identità dell'issuer riporta due spunte perché è stato usato il metodo interno, che però non è decentralizzato e fa affidamento su una lista centralizzata. Legato a questo, nelle conclusioni direi che se uno adotta la contromisura di usare una lista centralizzata, tanto vale evitare di usare la blockchain e fare affidamento su una PKI.}
As evident from Fig. \ref{fig:verification-normal}, the issuer identity verification results in two confirmations. While the second one relates to the successful decentralized verification of the (supposed) issuer domain, the first confirmation highlights that the issuer has been internally verified by Block.co, which is acting as a central trusted authority.

\section{A forgery attack on the Block.co protocol
\label{sec:forgery}
}

Let us suppose that an attacker aims at forging a Block.co certificate. To do this, they aim at impersonating a valid issuer and replicating all the steps of the issuance procedure to obtain certified credentials. 
This can indeed be achieved by exploiting the option of external issuer identity verification provided by the Block.co platform.
For this, it is sufficient that the attacker is provided with a valid Bitcoin address and sets up a public URL referring to the Bitcoin address used for the attack.
This can be easily done by manipulating the \textit{Issuer} metadata as shown in Fig. \ref{fig:mal-issuer} to enable the external issuer identity verification with the malicious Bitcoin address and the corresponding public URL.
For better clarity, the attack procedure is described in Algorithm \ref{alg:attack}.

\begin{figure}[htbp]
\centering
{\scriptsize
\begin{lrbox}{\jsonbox}
\begin{minipage}{0.8\linewidth}
\begin{verbatim}
{ 
    "name": "Name of the impersonated issuer",
    "identity": {
        "address": "Malicious Bitcoin address",
        "verification":[
            {"block_co": "false"}, 
            {"domain": "Malicious URL"}
        ]
    }
}
\end{verbatim}
\end{minipage}
\end{lrbox}

\fcolorbox{gray}{gray!10}{\usebox{\jsonbox}}
}
\caption{Malicious Issuer metadata}
\label{fig:mal-issuer}
\end{figure}

\begin{algorithm}
\caption{Block.co certificate forgery attack}\label{alg:attack}
\begin{algorithmic}[1]
\REQUIRE An owned Bitcoin address, a forged URL exposing the owned Bitcoin address.
\STATE Fill the \textit{Issuer} metadata with the forged URL and the Bitcoin address.
\STATE Craft the PDF certificate and insert the \textit{Issuer} metadata.
\STATE Build a Merkle tree with the certificate as a leaf.
\STATE Compute the Merkle root and extract the Merkle proof.
\STATE Send a Bitcoin transaction with the Merkle root inside the OP\_RETURN field.
\STATE Add the transaction identifier, the Merkle root, and the Merkle proof to the certificate metadata.
\RETURN Certificate ready to be shared and verified.
\end{algorithmic}
\end{algorithm}

%To be clear, the field \textit{block\_co} can also be omitted as it would produce the same result.

%By setting the \textit{block\_co} metadata to \textit{false} it is possible to bypass the internal issuer verification. For the external issuer verification it is only necessary to host on the malicious URL the Bitcoin address used for the attack. 

\subsection{Threat Model}
\label{subsec:threat-model}
We consider an attacker who aims to produce a forged digital credential that the Block.co verifier accepts as legitimately issued by an institution possibly accredited within the Block.co ecosystem, despite having no actual relationship with that institution. The attacker has access to a Bitcoin wallet capable of issuing transactions (only mining fees are required for OP\_RETURN transactions), a publicly reachable HTTPS URL where arbitrary content can be hosted, and standard cryptographic primitives to craft a certified PDF file and manipulate its metadata fields.

We assume that the attacker can populate the \textit{Issuer} metadata of the PDF with arbitrary values, including setting \textit{block\_co} to false (or omitting it) to bypass the centralized issuer verification, and that the attacker controls both the wallet address and the public URL declared in the metadata. We further assume, based on our experimental observations, that the Block.co verifier processes the submitted metadata as authoritative when \textit{block\_co} is false or omitted, without cross-checking the claimed issuer name against an internal registry and without enforcing consistency between the declared URL and the claimed institution. The Bitcoin network accepts the attacker's OP\_RETURN transaction as a standard operation requiring no special privileges.

Out of scope are attacks that break cryptographic primitives, compromise the Block.co verifier infrastructure, obtain private keys of legitimate accredited issuers or target the Bitcoin network itself (e.g., 51\% attacks).

\subsection{Experimental demonstration of the attack}

\begin{figure}[htp]
\centering
\includegraphics[width=0.95\linewidth]{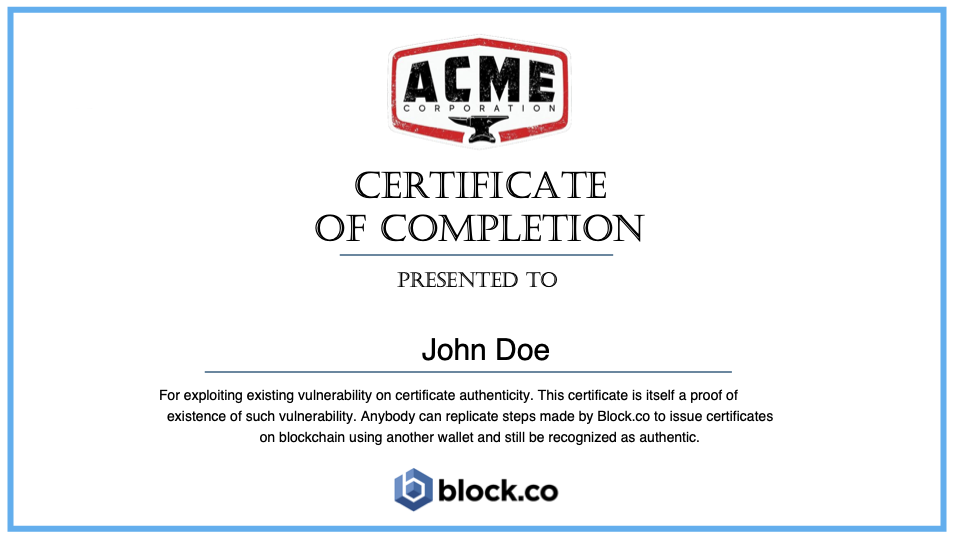}
\caption{Crafted PDF}
\label{fig:pdf-view}
\end{figure}

In order to demonstrate the feasibility of the attack, we crafted a certificate as falsely issued by Università Politecnica delle Marche.
This has been done by using a valid Bitcoin address, which is exposed on a public GitHub repository, and clearly does not correspond to any official account of our university.
Fig \ref{fig:pdf-view} shows the content of the certificate, which is falsely attesting the completion of some academic course and the achievement of the corresponding degree.

%{imgs/final_certified.png}

The malicious PDF metadata are populated with the information related to the forged certificate, as shown in Fig. \ref{fig:meta-mal-pre-cert}. %Standard metadata is omitted for the sake of brevity.

\begin{figure}[h!]
\centering
{\scriptsize
\begin{lrbox}{\jsonbox}
\begin{minipage}{0.95\linewidth}
\begin{verbatim}
Issuer = { 
   "name": "Universita Politecnica delle Marche",
   "identity": {
      "address": "bc1q8gwt3huv3md9qwtcx5r2atdq6jjll
                    ftfdv5tc5",
      "verification":[ 
         {"domain": "https://raw.githubusercontent.com/
                             giacomozonneveld/test/refs/
                             heads/main/"}
      ]
   }
}

Chainpoint Proof = {}
\end{verbatim}
\end{minipage}
\end{lrbox}

\fcolorbox{gray}{gray!10}{\usebox{\jsonbox}}
}

\caption{Malicious metadata before the transaction on blockchain}
\label{fig:meta-mal-pre-cert}
\end{figure}

Next, we compute the hash digest of such a malicious PDF file and we use it to create a Merkle tree with 10 leaves, where each leaf has the same hash digest. From such a Merkle tree, we extract the Merkle root and the Merkle proof of the first leaf, corresponding to the forged certificate.
Then, a Bitcoin transaction containing the computed Merkle root in the OP\_RETURN field is issued to the Bitcoin network from the malicious Bitcoin address. 
Clearly, such a transaction, which does not imply any cryptocurrency transfer, is acceptable by the Bitcoin network, provided that the malicious wallet is able to cover the corresponding mining costs.
Once the transaction has been accepted by the network and inserted into the Bitcoin blockchain, we complete the forgery procedure by populating the \textit{Chainpoint Proof} metadata of the forged file with all the required information, that is, the hash digest of the PDF file (\textit{targetHash}), the Merkle root (\textit{merkleRoot}), the Merkle proof (\textit{proof}) and the Bitcoin transaction identifier (\textit{anchors}), as shown in Fig \ref{fig:meta-mal-cert}.

\begin{figure}[h!]
\centering
{\scriptsize
\begin{lrbox}{\jsonbox}
\begin{minipage}{0.9\linewidth}
\begin{verbatim}
Chainpoint Proof = {
  "@context": "https://w3id.org/chainpoint/v2",
  "type": "ChainpointSHA256v2",
  "targetHash": "b4ffac2fc571228ab01a3fcd3c1a416643c
                    341951038a73855970bdf8e884f4e",
  "merkleRoot": "68d76584952eb372cc8ebd997eb23483b32
                    a5e0f51a841a5b225b4bdf9123e28",
  "proof": [
    {"right": "b4ffac2fc571228ab01a3fcd3c1a416643c34
                    1951038a73855970bdf8e884f4e"}, 
    {"right": "c6ff6ac078943b6efd483d9c4b555628d50e3
                    7e173fe7a5c275907a623620b89"}, 
    {"right": "9a288ce15259de3054d8a28ee315cd3a94476
                    244907a8d57eb6e8d750353c248"}, 
    {"right": "c6ff6ac078943b6efd483d9c4b555628d50e3
                    7e173fe7a5c275907a623620b89"}
  ],
  "anchors": [{
      "type": "BTCOpReturn",
      "sourceId": "2defd915263143398f4459034fcbcfd86b
                    aba30951c9cc39da85467cd42bf11f"
    }
  ]
}
\end{verbatim}
\end{minipage}
\end{lrbox}

\fcolorbox{gray}{gray!10}{\usebox{\jsonbox}}
}

\caption{Malicious Chainpoint Proof metadata}
\label{fig:meta-mal-cert}
\end{figure}

At this point, the malicious PDF file is forged and ready to be tested through verification.

\subsection{Verification}

\begin{figure}[htp]
\centering
  \includegraphics[width=0.95\linewidth]{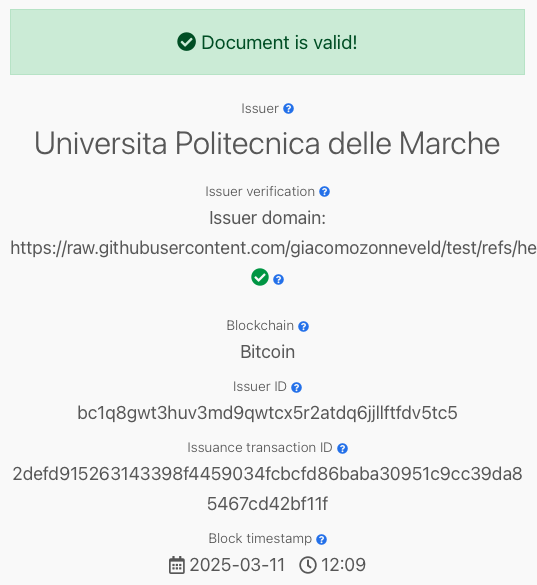}
\caption{Outcome of the verification of a forged PDF file}
\label{fig:verification-forged}
\end{figure}

%We submit the malicious PDF for its verification. By analyzing the behavior of the platform we conclude that the internal issuer verification is bypassed, so the credentials and the issuer are considered verified.
We can use the Block.co website for verifying the validity of the forged certificate.
By uploading our forged certificate, we obtain a positive verdict on both the document integrity and the issuer authenticity, as shown in Fig. \ref{fig:verification-forged}.

%\textcolor{red}{Qui aggiungerei due righe per spiegare che c'è una spunta sola perché stiamo usando la verifica esterna, che in effetti è l'unica sensata in quanto decentralizzata}
As expected, the issuer identity is being verified only in the decentralized manner, resulting in a single confirmation, compared to the verification of the legitimate certificate in Fig. \ref{fig:verification-normal}.
This type of issuer authentication, however, is the only one actually leveraging decentralization of the blockchain technology. In fact, the internal issuer verification through a list is centralized and could be replaced by a more robust centralized authentication based on a trusted identity provider and a PKI.

\section{Possible countermeasures
\label{sec:countermeasures}
}

\begin{comment}  
\begin{tcolorbox}[colback=blue!5!white, colframe=blue!75!black]
\begin{itemize}
    \item verifichi tutti gli issuer (e non accetti documenti da issuer non verificati)
    \item warning issuer
    \item pagina di pubblico dominio con tutti gli address degli issuer 
    \item blockchain permissioned (scrivono solo gli issuer abilitati da block.co e leggono tutti)
\end{itemize}
\end{tcolorbox}
\end{comment}

The vulnerability that has been identified is due to the fact that the process of decentralized verification of the issuer identity has no trust anchor and, thus, is circular within the decentralized blockchain infrastructure. Therefore, any user in the network can craft the same digital items as the real issuer and claim to be the real issuer, without the real issuer being either involved or made aware in any way.
This makes it easy to carry out a forgery attack.
In order to circumvent such a vulnerability, suitable countermeasures are needed. The primary concern pertains to the challenge of validating issuers in a decentralized manner. 

A first solution would consist in showing a clear alert on the issuer verification section when the issuer is not verified by some authority (like Block.co itself or, even better, a trusted identity provider).
%Accordingly, if the decentralized issuers validation can not be achieved, an internal database containing verified issuers should be maintained. This database could then be consulted when the PDF is submitted for validation. In the event that the issuer is not present in the database, the credentials should be refused. 
This solution would increase security, but sacrificing decentralization. 
In fact, in this case, the use of blockchain technology could be avoided, and replaced by a more traditional PKI.
An additional measure that could be implemented consists in maintaining a publicly accessible website listing all verified issuers and their associated blockchain addresses. This way, users can directly verify that the blockchain address from which transactions are originated is legitimate.
Again, however, the solution would be at the expense of decentralization, and the problem would shift to the need for robustness of the centralized list, which would again require measures similar to those needed to implement a centralized trusted identity provider.

A different and (though partially) decentralized solution would be to use a permissioned blockchain, wherein only issuers verified by a consortium of network nodes having greater privileges than the others are allowed to write transactions containing certification proofs of credentials, while any other network node is able to read such transactions to verify the validity of a certificate.

A fully decentralized solution, on the other hand, appears difficult to identify. One possibility in this regard is the use of DIDs\footnote{\url{https://github.com/uport-project/ethr-did-registry}}, which join the decentralization of blockchain with the need to use centralized authentication mechanisms. In that case, each issuer could be authenticated through different types of credentials (although each of them remains centralized), for example issued by the relevant national authorities, but released in the form of verifiable credentials.
In particular, some solutions based on anonymous credentials and zero-knowledge proofs \cite{Mosakheil2024} appear suitable for creating a decentralized authentication system that is not prone to forgery attacks of the type illustrated in this paper.

In conclusion, fully decentralized credential management and certification is infeasible, since it is mandatory to create a trusted relationship between the real identity and the digital identity of the issuing institution.

%\par \textcolor{red}{Soluzione fake me \cite{fakeme} possibile ma richiede collaborazione di molti enti per cui difficile da realizzare. Soluzione \cite{Serranito2020} buona, ma secondo smart contract richiede continue aggiunte di nuovi certificati creati oltre ad avere un'autorità centrale per la gestione del registro contenente i did document. Una possibile soluzione potrebbe essere un ibrido tra \cite{Serranito2020} e uso di merkle tree. Potrebbe esserci uno smart contract per le varie HEI per coprire la parte di autenticità (con autorità che certifica le identità dei vari hei) e continuare a usare merkle tree e transazioni per integrità... il tutto su public blockchain come ethereum}

\section{Responsible Disclosure}
\label{sec:disclosure}

We followed responsible disclosure practices before publishing this work. The vulnerability described in this paper was reported to Block.co by opening ticket BC-1344, entitled ``Vulnerability of Block.co to certificate forgery'', created on 6 May 2025. %\st{No response, acknowledgment, or remediation timeline has been received to date.} 
On 20 May 2026, we re-tested the originally crafted forged certificate against the public verifier, confirming that the vulnerability remains exploitable. On 3 June 2026, we reported the vulnerability to Block.co again. Block.co confirmed the vulnerability, temporarily suspended the certified document verification service, and is reviewing it.

\section{Conclusion}
\label{sec:concl}

The adoption of blockchain technology to certify digital credentials leads to significant advantages, but also to new attack surfaces.
The proposed analysis demonstrates a vulnerability in the Block.co protocol, which allows the creation of fake certificates through the impersonation of unverified issuers. The attack exploits the weakness of decentralized verification mechanisms, which are based on public URLs and Bitcoin addresses. To mitigate this risk, we proposed some possible countermeasures, involving the adoption of permissioned blockchains or more robust issuer verification mechanisms, even if they come at the price of some centralization of the underlying infrastructure.

\bibliographystyle{IEEEtran}
\bibliography{biblio}

\end{document}